\documentclass[aps,showpacs,twocolumn,pre,showpacs,color]{revtex4}

\usepackage{amsmath}
\usepackage{color}
\usepackage{amsfonts}
\usepackage{times}
\usepackage{graphicx}

\usepackage{amssymb}
\usepackage{calc}

\pdfoutput=1

\begin{document}

\title{Ordering dynamics of blue phases entails kinetic stabilization of amorphous networks}


\author{O. Henrich$^1$,  K. Stratford$^2$, D. Marenduzzo$^1$, M. E. Cates$^1$}

\affiliation{$^1$ SUPA, School of Physics and Astronomy, University of Edinburgh, Mayfield Road, 
Edinburgh EH9 3JZ, UK \\
$^2$ EPCC, The University of Edinburgh, Mayfield Road,
Edinburgh EH9 3JZ, UK}







\begin{abstract} The cubic blue phases of liquid crystals  are fascinating and technologically promising examples of hierarchically structured soft materials, comprising  ordered networks of defect lines (disclinations) within a liquid crystalline matrix. We present the first large-scale simulations of their domain growth, starting from a blue phase nucleus within a supercooled isotropic or cholesteric background. The nucleated phase is thermodynamically stable; one expects its slow orderly growth, creating a bulk cubic. Instead, we find that the strong propensity to form disclinations drives the rapid disorderly growth of a metastable amorphous defect network. During this process the original nucleus is destroyed; re-emergence of the stable phase may therefore require a second nucleation step. Our findings suggest that blue phases exhibit hierarchical behavior in their ordering dynamics, to match that in their structure.
\end{abstract}

\maketitle

The blue phases, BPI and BPII, of chiral nematic liquid crystals can each be viewed as an ordered network of topological defects (disclination lines), embedded within a liquid crystalline matrix (a cholesteric) whose local molecular alignment axis rotates helically on moving through the sample. Such phases were long viewed as purely of academic interest, due to their extremely narrow window of thermodynamic stability (of order $1$K) \cite{blue1}. This has changed completely with the recent development of new compounds showing stable BPs over a $50$K interval, with fast switching between different states \cite{blue2,blue3}. Thus BPs now offer a promising device technology not only for displays \cite{blue4} but also for laser applications \cite{blue5,blue6}. To fully realize this potential requires an understanding of how chiral nematic materials switch between different structures. However, theoretical work on BPs has advanced relatively modestly since the late 1980s \cite{blue1}, and so far there is almost no understanding of their phase-change kinetics. This is partly because of computational challenges which, despite pioneering progress using small systems \cite{yeomans1,yeomans2,yeomans3}, have prevented the simulation of supra-unit cell behavior in the ordered BPI and BPII, and ruled out realistic simulation of a third blue phase, the apparently amorphous \cite{blue1} BPIII. 

With the aid of supercomputers and a hybrid lattice Boltzmann algorithm \cite{marendu1}  we have recently overcome these difficulties, enabling us to address systems hundreds of times larger than the unit cell volume. This has allowed us to address elsewhere the motion of planar interfaces between competing phases \cite{marendu2}. By the same methods, we address below for the first time the domain growth of the ordered BPs from an isolated nucleus of the stable phase. We find that the growth process is unexpectedly interrupted by the proliferation of a metastable BPIII-like structure. Thus the system seemingly opts for speed rather than efficiency, lowering its free energy rapidly by amorphous defect proliferation even though this creates barriers that prevent attainment of the global free energy minimum -- which could have been reached directly by a slower, orderly growth of the initial nucleus. 
This finding, which may have wide implications for phase kinetics in other multi-scale soft materials (see e.g. \cite{multiscale,multiscale2}), is sharply different from some other instances where metastable phases intervene in phase ordering (`Ostwald's rule') as we discuss below.

Experimentally BPs are found adjacent to the cholesteric phase which has local nematic orientational order, along a director that slowly rotates in space; in the cholesteric, there is a single rotation axis and the wavelength in this direction is called the pitch. (All these phases lie below the isotropic phase in temperature; see  Fig.1 for a phase diagram.) Unlike the cholesteric, the BPs have director modulation in all directions and comprise an interpenetrating lattice (BPI/II) or disordered network (BPIII) of so-called `double twist cylinders' \cite{blue1}. These accommodate a higher degree of director twist than the simple cholesteric, at the price of creating a complementary network of topological defects (disclination lines) on which the nematic ordering vanishes; the resulting defect-lattice unit cells in BP I/II are shown in Fig.1. The blue phases are more stable than the simple cholesteric at high enough molecular chirality. 

It is known that, on the scale set by the isotropic-cholesteric transition, the free energy differences among BPs  are relatively small. Nonetheless, per unit cell, these differences typically remain large compared to the thermal energy, $k_BT$ \cite{blue1}. This suggests only a limited role for thermal noise in BPs (but see \cite{stark} for a contrary view). Thus one might expect strong hysteresis and metastability of phases, but this has not been reported experimentally \cite{blue1,blue2,blue3,phasework}. It is possible \cite{smectics} that small nuclei of BPs persist on heating far into the isotropic phase, allowing hysteresis-free growth on reducing the temperature again. Whatever the source of the required nuclei, simulation studies of their subsequent growth can illuminate fundamental issues in phase transition dynamics, as exemplified by previous work on hard-sphere colloids \cite{auer1,auer2}.

Our hybrid lattice Boltzmann technique is summarized in the methods section and in supporting information (SI; Text S1), and detailed elsewhere \cite{marendu1,marendu2,marendu3}.
Briefly, it marries a finite difference code for the advective relaxation of the order parameter tensor ${\bf Q}({\bf r})$ (with ${\bf r}$ spatial position) \cite{beris}, governed by a suitable free energy functional ${\cal F}[{\bf Q}]$ \cite{deGennes}, to an efficient parallel lattice Boltzmann (LB) code for a forced Navier-Stokes equation for momentum transport, like that used previously to simulate multiscale colloidal materials \cite{kevin}.

We have found in previous simulations \cite{marendu2} that a semi-infinite slab of stable BPI or BPII will invade an adjacent region of supercooled isotropic or cholesteric phase. In the present work we extend these studies to the more realistic case of a localized nucleus, comprising a few unit cells (typically 8) of the stable BP. This nucleus is prepared by excising it numerically from a periodic BPI or II, placing it within the chosen environment, and then allowing it to anneal (see Methods section). Annealing is done very close to the coexistence boundary between phases, where there is no tendency for the nucleus to grow; during annealing its surface reconstructs to achieve local equilibrium with the surrounding phase. We then quench the system, altering the reduced temperature $\tau$ and/or dimensionless chirality $\kappa$ (defined in supporting Text S1) so that the nucleated BP is now the equilibrium structure. In most of our simulations there is no thermal noise; adding modest amounts of noise makes quantitative but not qualitative differences to our findings (with one exception detailed below).

Fig.2(a-c) shows the time evolution of a BPII seed in an isotropic environment. Depicted is an `isosurface' at a fixed small magnitude of nematic order (defined through the largest eigenvalue $0<q<1$ of the order parameter tensor ${\bf Q}$; see methods section). In a bulk BP this representation allows unambiguous imaging of the disclinations, each of which is wrapped by a thin tube of the isosurface. Here, since there is no nematic order in the external phase, this isosurface envelops the growing nucleus; to expose the structure within, we omit in Fig.2b the front section of the growing droplet. Although there is residual cubic anisotropy (smaller with thermal noise than without) the phase formed does not have the long range order of BPII; it is instead an amorphous, aperiodic network. This disorder is also visible in the structure factor $C({\bf k})$ \cite{supporting} shown in Fig.2d. By the end of the run, no trace remains of the initial nucleus.

Fig.3 shows a similar sequences for a BPI seed in a cholesteric matrix.  The pattern of evolution is similar although
the residual anisotropy is no longer cubic (Fig.3d), as befits the lowered initial symmetry. The dynamics of the defect proliferation can be clarified further by using instead an extended rod-like nucleus of square cross section, whose long axis spans the periodic simulation box. When viewed along that axis (Fig.4) the noise-free growth of the disclination pattern, though aperiodic, retains the symmetry of the initial state. Addition of thermal noise allows this symmetry to break and increases the rate at which order is lost (see supporting Fig.S1); this is the expected consequence of adding weak noise to a time evolution governed by deterministic, but probably chaotic, dynamics.
Finally, in Fig.5 we show growth of a BPI seed in an isotropic environment. In this case, the quench parameters are the same as in Fig.2, so that the nucleated phase BPI is metastable, with BPII stable. As before, an amorphous network forms in preference to either ordered phase. Supporting Videos S1-S4 show the complete time evolutions corresponding to Figs.2-5.

These results demonstrate that seeding a supercooled isotropic or cholesteric phase with a small BPI or BPII nucleus leads to growth, not of the stable phase, but of metastable, amorphous networks. Although clearly BPIII-like, these states probably do not belong to a metastable branch of BPIII itself (see supporting Text S1). A more important question, especially given the observed reversibility of the experimental phase diagram, is whether such amorphs can thereafter become ordered. We find that, at least for low chirality ($\kappa\lesssim 2$), this transformation is not spontaneous: without noise, the system relaxes almost to a standstill, suggesting that a second nucleation is needed to reach equilibrium. We thus tried directly re-inserting a BPI/II nucleus similar to the initial one; this disappears, indicating that the critical nucleus for this step is relatively large.  At higher chirality, we see instead continuous ripening of the amorph towards of a long-range ordered phase (supporting Fig.S2). However, this is not BPI or II, but another phase ($O^5$) long predicted by theory \cite{blue1} but not yet seen experimentally. More generally, our findings admit a variety of kinetic pathways that might connect the metastable amorph formed initially to BPI/II, although in some cases this final equilibration might not be observable on any experimental timescale. 

\begin{figure*}[ht]
\vspace*{0.05in}
\includegraphics[width=0.6\columnwidth]{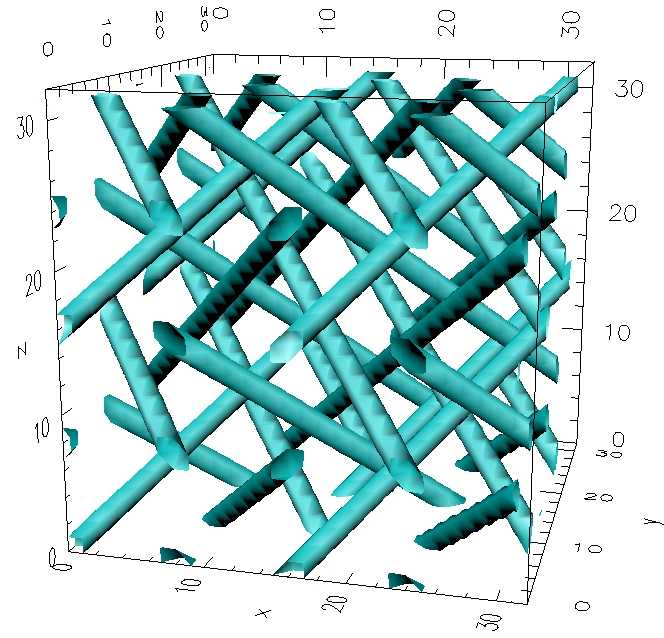}
\includegraphics[width=0.6\columnwidth]{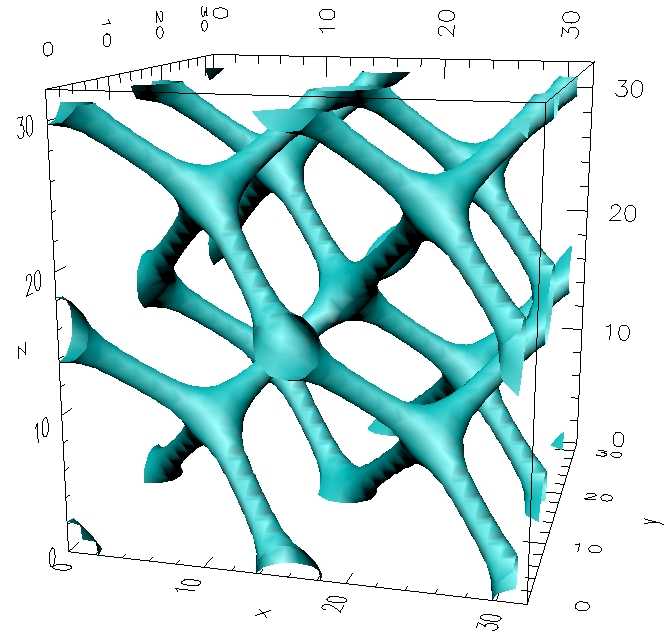}
\includegraphics[width=0.6\columnwidth]{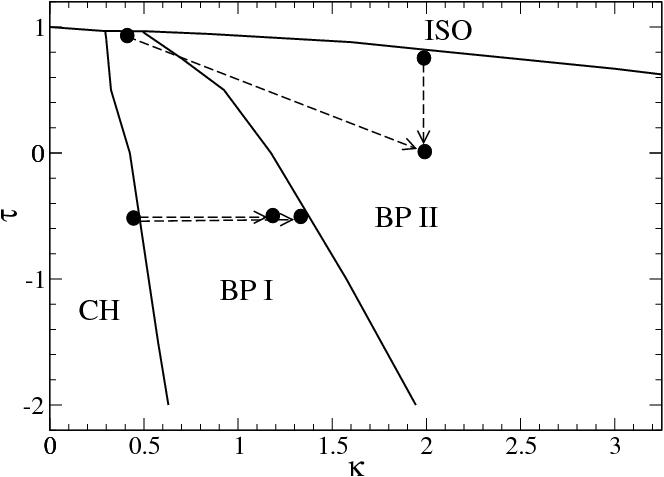}
\caption{Left and centre: disclination network of the crystalline Blue Phases BPI (left) and BPII (centre). The pictures show $8$ unit cells, $2$ along every coordinate direction. Right: computed phase diagram in the plane spanned by chirality ($\kappa$) and effective temperature ($\tau$). At high chiralities BPII becomes unstable relative to either the O5 structure (in theory) or the BPIII structure (in experiment); these phases are not shown here. The dotted arrows show the various quenches used in Figs.2-5, from states close to a phase boundary to the final parameters at which domain growth occurs.}
\end{figure*}

\begin{figure*}[ht]
\vspace*{0.05in}
\includegraphics[width=0.5\columnwidth]{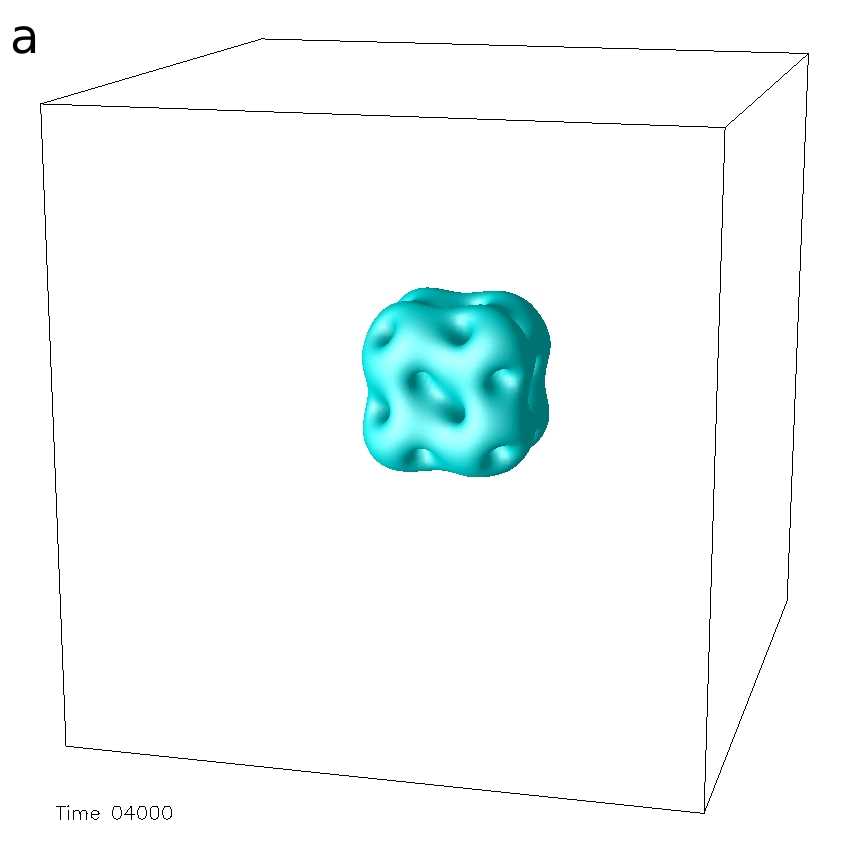}
\includegraphics[width=0.5\columnwidth]{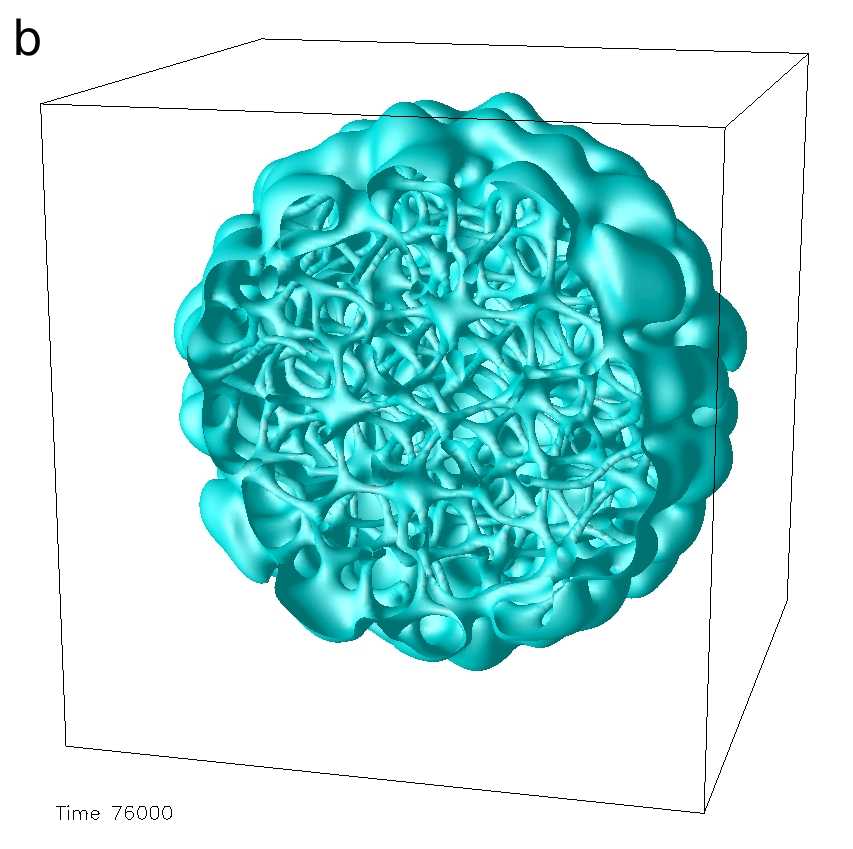}
\includegraphics[width=0.5\columnwidth]{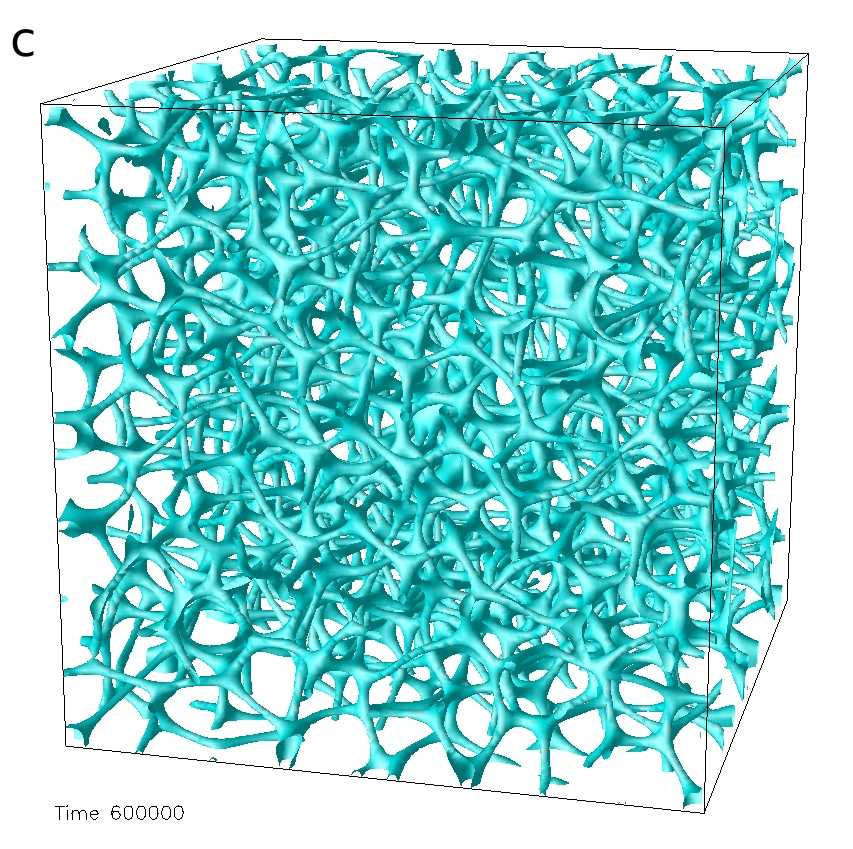}
\includegraphics[width=0.5\columnwidth]{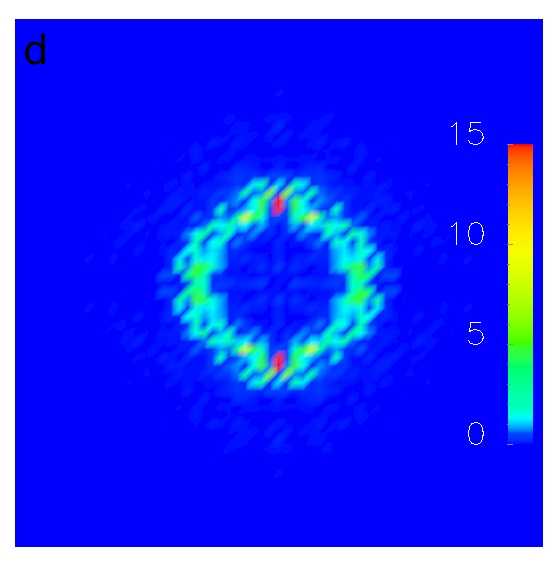}
\caption{Domain growth of a BPII droplet in isotropic environment. The droplet was equilibrated near the BPII-isotropic boundary and then quenched to $\tau=0, \kappa=2$, where BPII is the equilibrium phase. The pictures show isosurfaces ($q=0.13$) of the scalar order parameter during equilibration, and at intermediate and late times ($t=4\times 10^3$; $t= 7.6 \times 10^4$; $t=6\times 10^5$) (a-c). (For typical materials the simulation timestep corresponds to $\sim 1$ ns.) Frame (d) gives a cut through the structure factor of the final state at $k_z=0$ for wavevectors $k_x$ and $k_y$ in $[-\pi/2\ell,\pi/2\ell]$. Here $\ell$ is the discretization length of the LB lattice (about $10$nm for typical materials).}
\end{figure*}
\begin{figure*}[ht]
\vspace*{0.05in}
\includegraphics[width=0.5\columnwidth]{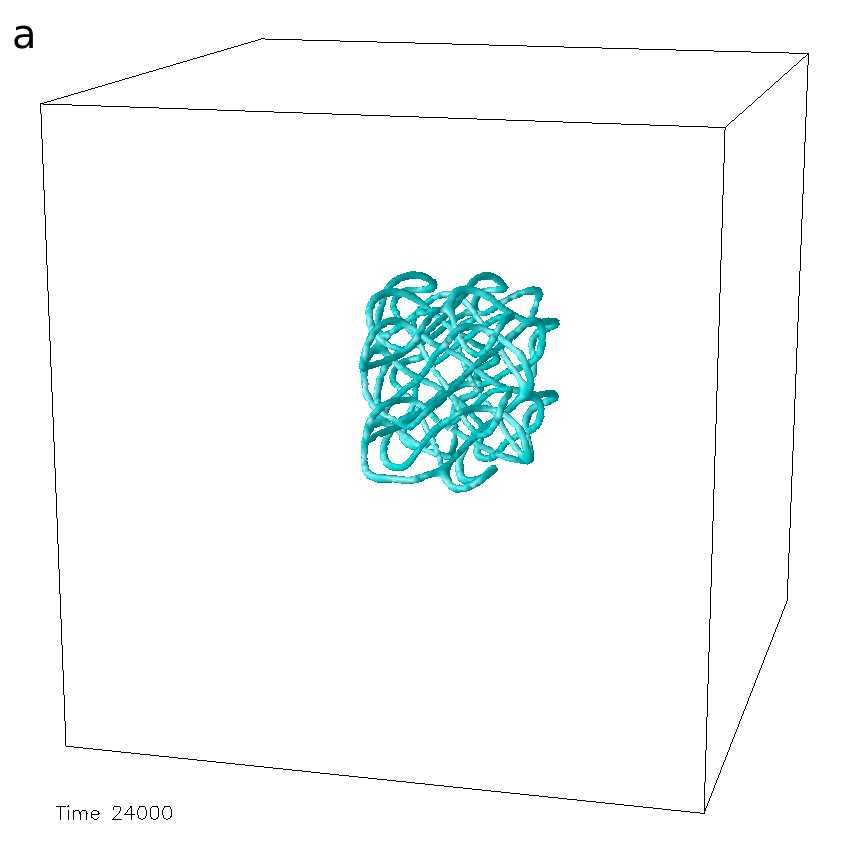}
\includegraphics[width=0.5\columnwidth]{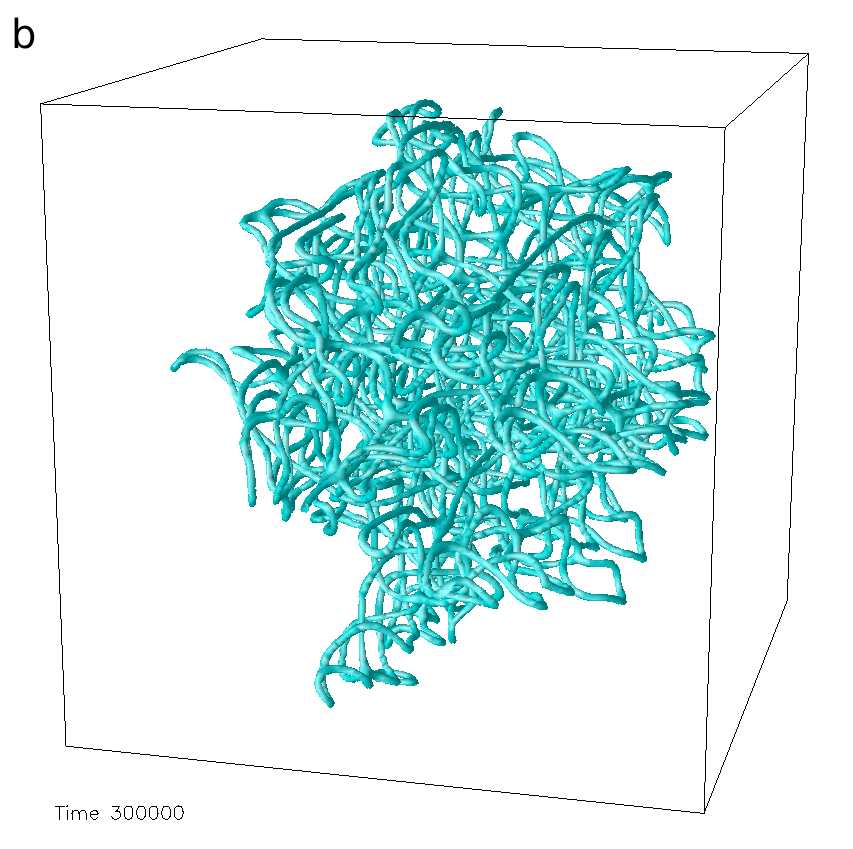}
\includegraphics[width=0.5\columnwidth]{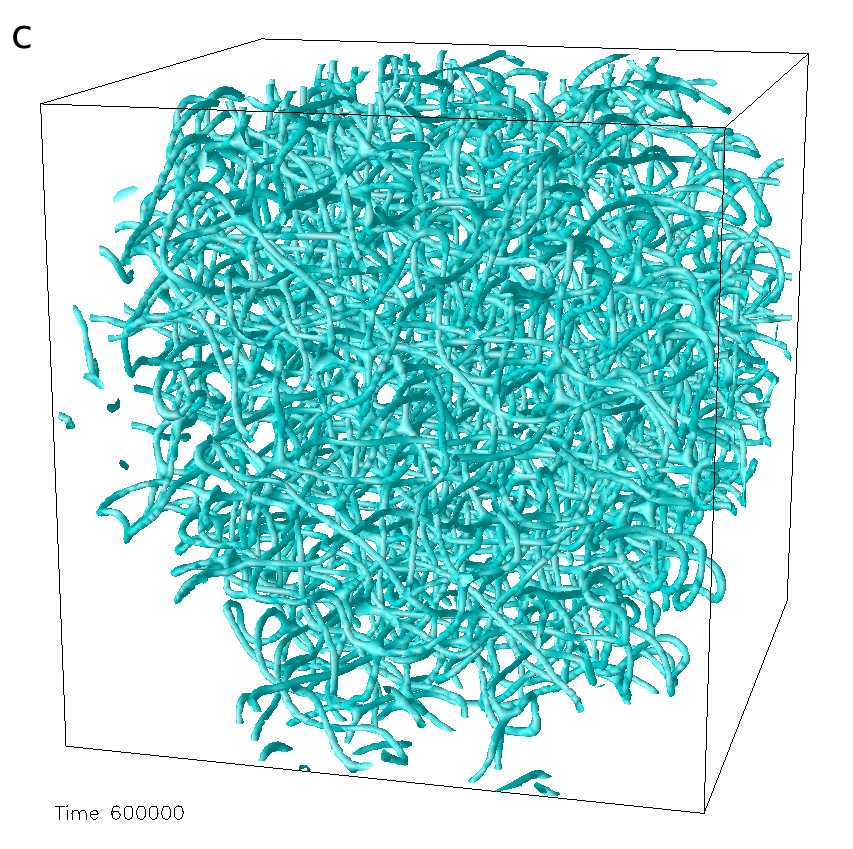}
\includegraphics[width=0.5\columnwidth]{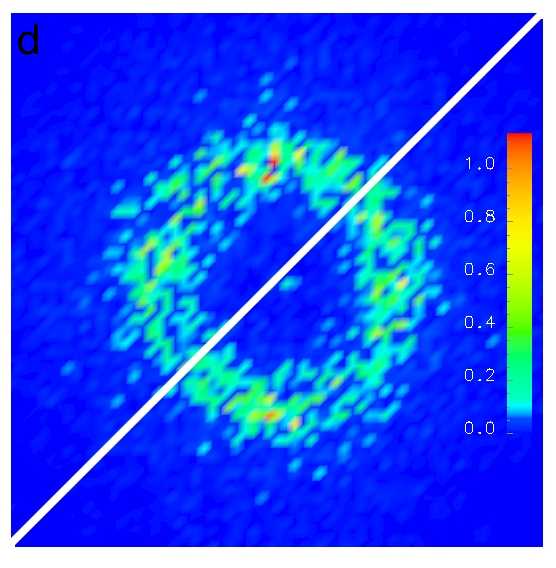}
\caption{Domain growth of a BPI droplet in cholestric environment. The droplet was equilibrated near the cholesteric-BPI boundary and then quenched to $\tau=-0.5, \kappa=1.2$, where BP I is the equilibrium phase. The pictures show isosurfaces ($q=0.2$) of the scalar order parameter immediately after the quench at $t=2.4 \times 10^4$ timesteps (a); and at two later stages, $t=5\times 10^5$ and $t=6\times 10^5$ (b and c). The top left half in (d) is a cut through the structure factor $C({\bf k})$ of the final state (c) along $k_y=0$ for $k_x\le k_z$ in the interval $[-\pi/2\ell,\pi/2\ell]$, whereas the bottom right half depicts a cut along at $k_z=0$ for $k_x\ge k_y$ in the same range.}
\end{figure*}

At first sight, our results are reminiscent of Ostwald's rule of stages \cite{ostwald}, whereby nucleation is asserted to proceed through a series of metastable phases, starting with the one whose free energy lies closest to (but below) that of the initial bulk. There is however no simple link: Ostwald's rule is normally attributed to having a higher barrier for nucleating the final structure directly \cite{frenkel}, whereas our simulations begin with a suitable nucleus already present. Moreover we have performed simulations with $\tau$ and/or $\kappa$ adjusted (supporting Tables S1, S2) to ensure that the initially nucleated ordered BP now lies closest in free energy to the initial isotropic phase. An amorphous network still forms, now with free energy lower than the nucleated phase, in contradiction to Ostwald's rule.

In conclusion, we have shown that the growth of a nucleus of an ordered BP within an isotropic or cholesteric matrix is generally interrupted by proliferation of a metastable amorph, despite the fact that this creates a barrier to formation of the equilibrium phase which could, however, have been created directly from the initial nucleus, whose structure it shares.
Our results illustrate the subtle character of domain growth in soft materials exhibiting hierarchical ordering in equilibrium. For BPs, the organizational levels are local cholesteric ordering; defect formation within the cholesteric matrix; and the long-range organization of the defects into a periodic lattice. Our results suggest a corresponding multi-stage time evolution, which can easily get stuck at an intermediate level in the hierarchy. Once the amorphous defect network has formed, the energetics is already optimized at the first two levels; the driving force for creating an ordered BPI/II superstructure is then relatively weak, and the barriers high. Conversely, the final thermodynamic preference for that ordered structure, even if a nucleus of it is initially present, is apparently swept aside in the rush to create disclinations during the earlier defect-proliferation stage. These findings may have wide implications for the phase ordering dynamics of multiscale soft materials --- increasingly relevant to nanotechnology, photonics, and display applications --- of which the blue phases offer striking and elegant examples.

\begin{figure*}[ht]
\vspace*{0.05in}
\includegraphics[width=0.5\columnwidth]{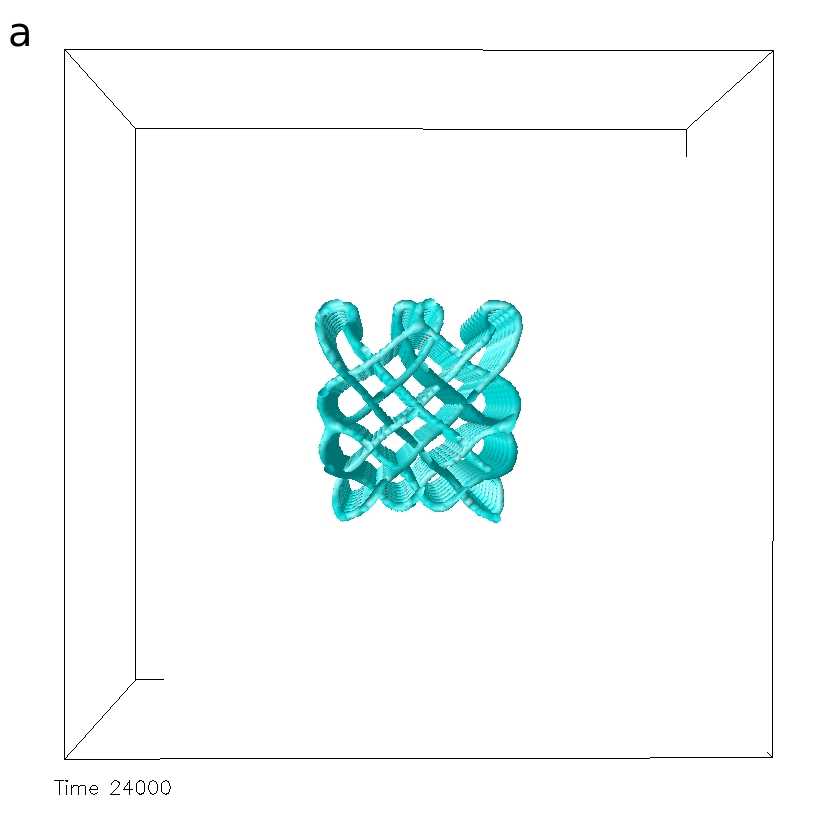}
\includegraphics[width=0.5\columnwidth]{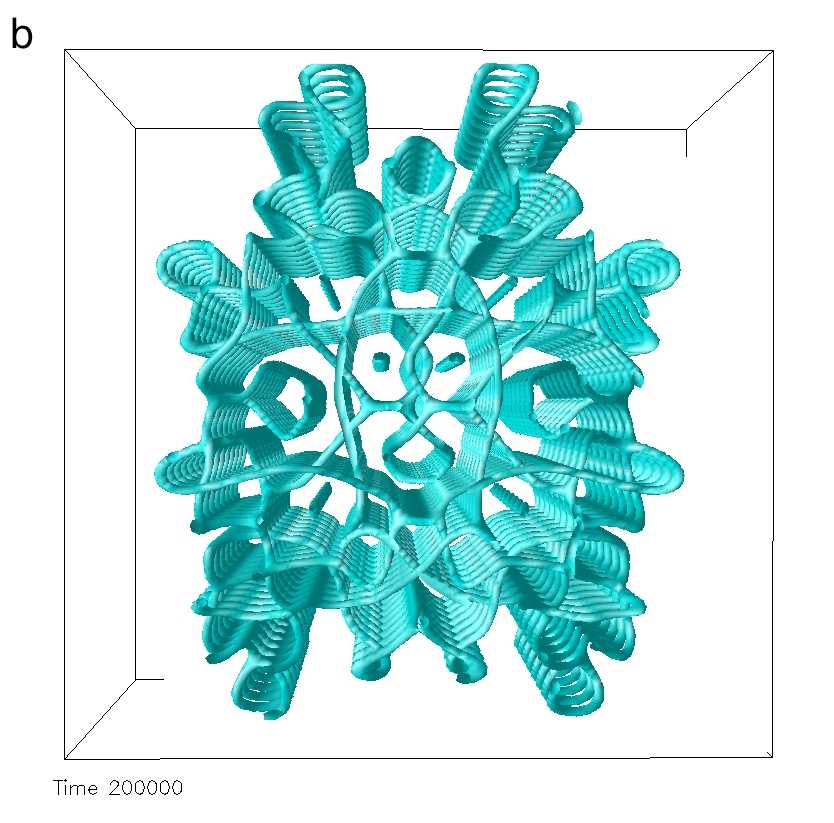}
\includegraphics[width=0.5\columnwidth]{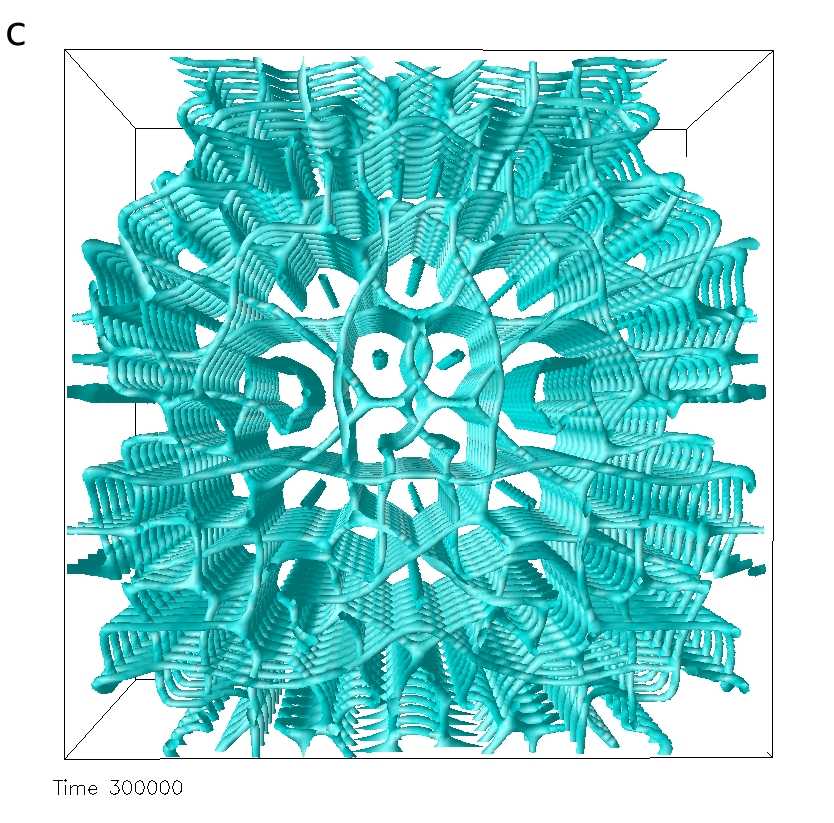}
\includegraphics[width=0.5\columnwidth]{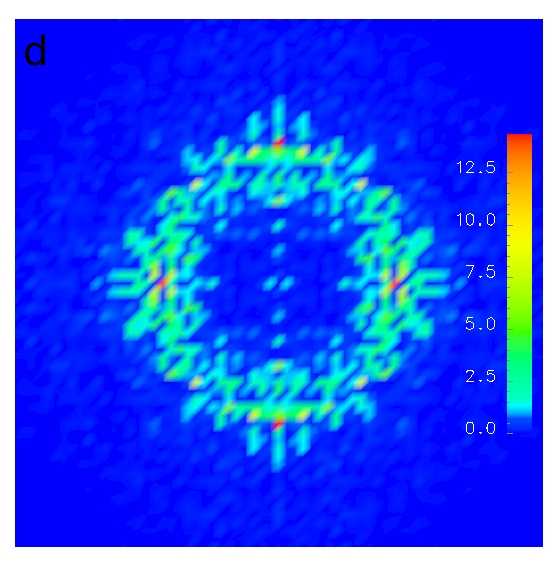}
\caption{Domain growth of a BPI rod in cholestric environment: The nucleus was equilibrated close to the BPI-isotropic boundary and then quenched to $\tau = -0.5,  \kappa = 1.35$. The pictures show isosurfaces ($q=0.2$) of the scalar order parameter immediately after the quench at $t=2.4\times 10^4$ timesteps (a), and at two later stages $t=2\times 10^5$ and $t=3\times 10^5$ (b and c).  (d): The structure factor of state (c) at $k_z=0$ for wavevectors $k_x$ (horizontal axis) and $k_z$ (vertical axis) in the interval $[-\pi/2\ell,\pi/2\ell]$.}
\end{figure*}
\begin{figure*}[ht]
\vspace*{0.05in}
\includegraphics[width=0.5\columnwidth]{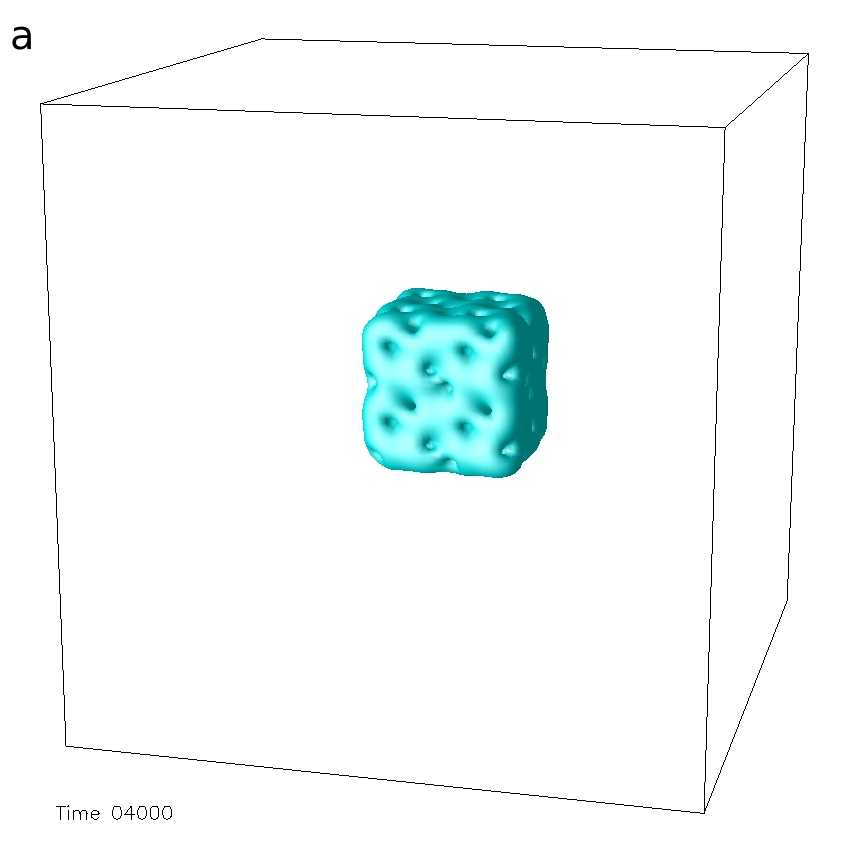}
\includegraphics[width=0.5\columnwidth]{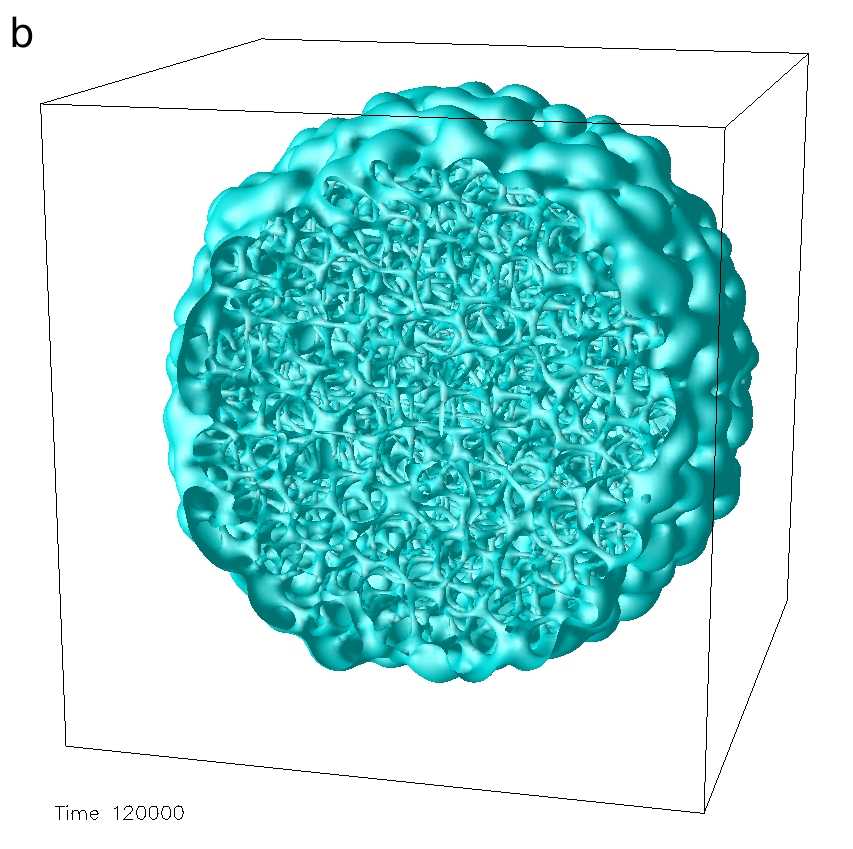}
\includegraphics[width=0.5\columnwidth]{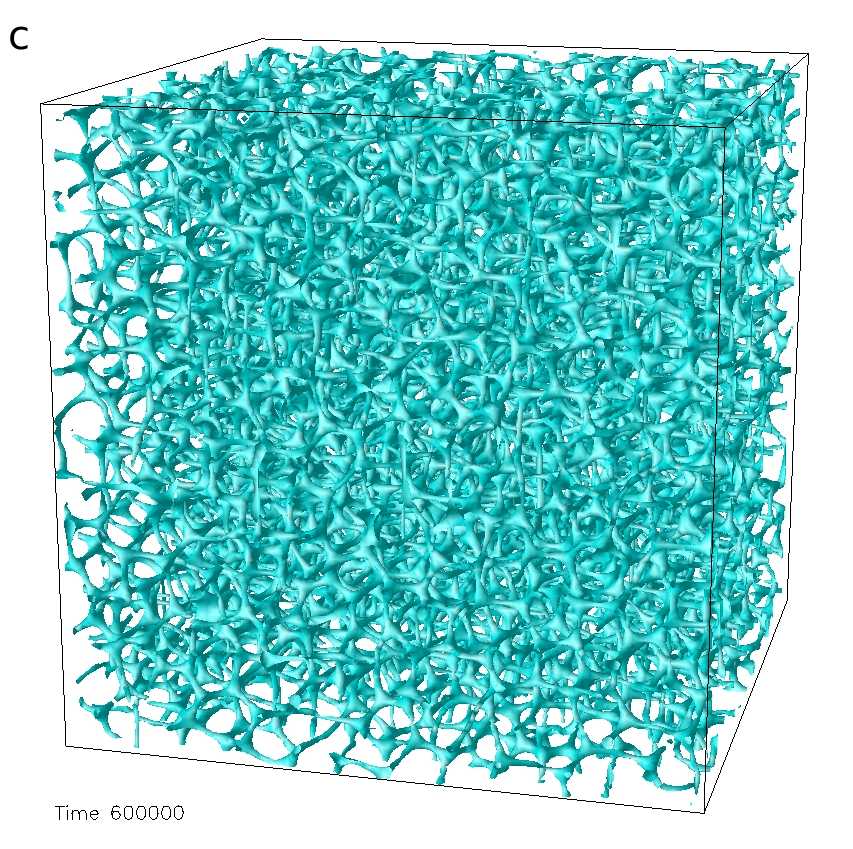}
\includegraphics[width=0.5\columnwidth]{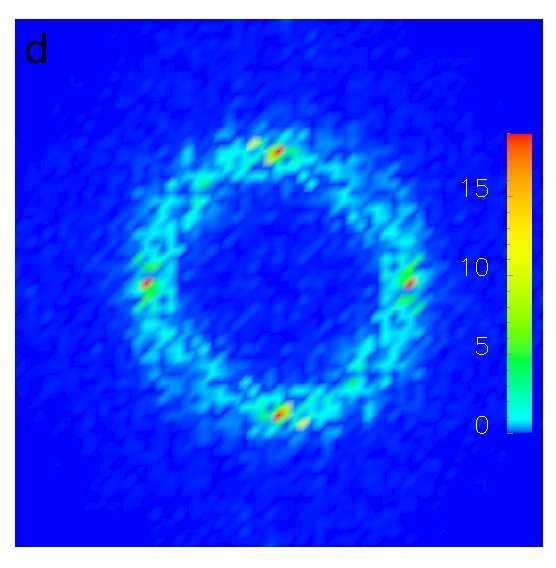}
\caption{Domain growth of a BPI droplet in an isotropic environment. The droplet was equilibtated close to the isotropic-BPI boundary and then quenched to $\tau=0, \kappa=2$, where BPII is stable. (See phase diagram in Fig.1.) The pictures show isosurfaces ($q=0.13$) of the scalar order parameter: during the equilibratio and at intermediate and late times (at $t=4\times 10^3$;$t=1.2\times 10^5$; $t=6\times 10^5$) (a-c). In picture (b) the growing domain has been cut at $y=32$ to reveal the internal structure. 
A cut at $k_z=0$ through the stucture factor $C({\bf k})$ of the final state is depicted in (d). The section shown corresponds to wavevectors $k_x$ (horizontal axis) and $k_y$ (vertical axis) in the interval $[-\pi/2\ell,\pi/2\ell]$.}
\end{figure*}

\section{Equations of Motion}
The free energy functional \cite{deGennes}, and the equations of motion \cite{beris} deriving from it, are presented in  supplementary Text S1. As detailed there, the control parameters of the free energy functional are conventionally chosen as $\tau$, a reduced temperature, and $\kappa$ a dimensionless chirality. We also report there our estimates of the free energy differences among BPs and specify the way in which thermal noise is introduced.

\section{Quench Details}
For each quench shown in Figs.2-5, we report in supplementary Table S2 the free energy densities for the stable BPI/II phases and for the metastable BPIII-like end state. 
Every simulation comprised two separate parts, namely the creation and equilibration of the initial configuration, followed by a quench in temperature and/or chirality. In order to generate an equilibrated nucleus, we started from the analytically known structures \cite{blue1} that describe the infinite chirality limits of BPI and BPII and first evolved these for 2000 time steps at the chosen initial $\kappa$ and $\tau$. This creates an equilibrium bulk phase at those parameters.
Then we replaced the tensor order parameter at all sites outside of a cube comprising 8 BP unit cells, situated at the centre of the simulation box, with either a cholesteric or an isotropic configuration.
This creates a nucleus of BP crystal within the required phase, but the nucleus has sharply-cut surfaces that could have unrealistically high surface energy, perturbing the subsequent dynamics. This state was therefore equilibrated in a second stage for another 18000 time steps, maintaining $\tau$ and $\kappa$ at the chosen initial values, which lie just within the BP equilibrium region but very close to the coexistence line with the surrounding phase. By this process we allowed the surface structure and shape of the BP nucleus to equilibrate locally, without initiating significant domain growth.
After the equilibration was finished, we performed a sudden quench to parameter values further away from the phase boundary, thereby initiating the domain growth. Note that for technical reasons (the requirement of fitting a whole number of unit cells into the simulated domain), when nucleation is done within a cholesteric matrix, the cholesteric pitch, post-quench, is not the optimal one. The same would apply in generic laboratory conditions involving samples of fixed geometry where a pure temperature quench will typically alter $q_0$ as well as both $\tau$ and $\kappa$. 

\section{Structural Visualization}
The openDX software package was applied to the simulated $q({\bf r})$ data to create the constant-$q$ isosurfaces shown in Figs.1-5. Choosing a small finite $q$ renders the disclination lines as tubes without topological ambiguity; the visually optimal choice of $q$ depends on the structure being viewed. The structure factor $C({\bf k})$ is defined as $|q({\bf k})|^2$ with $q({\bf k})$ the Fourier-transform of $q({\bf r})$ computed using routines from the FFTW library. 

\section{Computational Resources} A typical computational run time was 5-10 hours on 512 quad-core nodes of an IBM Blue Gene/P system, using a LB lattice of $128^3$ sites capable of containing around 500 BPI/II unit cells. For typical cholesteric materials, the LB lattice spacing corresponds to about 10 nm, and the timestep about 1 ns (see supplementary Text S1). Similar parameter mappings arise in LB simulations of hierarchical colloidal materials \cite{kevin}.

\begin{acknowledgments}
We acknowledge EPSRC (Grants EP/ E045316, EP/ E030173, EP/ F054750 and EP/ C536452) for funding and supercomputer time. We also acknowledge a DOE INCITE 2009 grant entitled `Large scale condensed matter and fluid dynamics simulations' at Argonne Leadership Computing Facility, supported by the US Department of Energy under contract DE-AC02-06CH11357. MEC holds a Royal Society Research Professorship.
\end{acknowledgments}

\pagebreak
\appendix

\section{Supporting Text}

Here we provide further information on the free energy functional and the control parameters $\tau, \kappa$ (section 1); on the equations of motion (section 2); on the conversion of simulation parameters to physical units (section 3); on the introduction of thermal noise (section 4); on the parameter values for our quenches (section 5); and on the end-state free energies in comparison with various stable and metastable phases (section 6).

\subsection{Free energy functional, chirality and reduced temperature}

The thermodynamics of cholesteric blue phases can be described via a 
Landau-de Gennes free energy functional $\cal F$, which in turn is
an integral over space of a free energy density $f$,
\begin{eqnarray}
{\cal F}[{\bf Q}]&=&\int d^3{\bf r} f({\bf Q}({\bf r})).
\end{eqnarray}
The free energy density $f=f({\bf Q})$ may be expanded in powers of the order parameter ${\bf Q}$ and its gradients; ${\bf Q}$ is a traceless 
and symmetric tensor.
The largest eigenvalue $q$ and corresponding eigendirection ($\bf n$) respectively
describe the local strength and major orientation axis of molecular order.
The ${\bf Q}$ tensor theory, rather than a theory based solely on the director field ${\bf n}({\bf r})$, allows treatment of disclinations (defect lines) in whose cores ${\bf n}$ is undefined; in blue phases, disclinations organise into regular or amorphous networks.

The free energy density we use, following \cite{blue1}, is:
\begin{eqnarray}
f({\bf Q})&=&\frac{A_0}{2}\left(1-\frac{\gamma}{3}\right)Q^2_{\alpha \beta}\nonumber\\ &-&\frac{A_0\gamma}{3}Q_{\alpha \beta} Q_{\beta \gamma}Q_{\gamma \alpha} + \frac{A_0\gamma}{4} (Q^2_{\alpha \beta})^2  \nonumber\\
&+&\frac{K}{2}(\varepsilon_{\alpha \gamma \delta} \partial_\gamma Q_{\delta \beta} + 2 q_0 Q_{\alpha \beta})^2+ \frac{K}{2}(\partial_\beta Q_{\alpha \beta})^2 \label{free}
\end{eqnarray}
Here repeated indices are summed over and $Q^2_{\alpha\beta}$ stands for $Q_{\alpha\beta}Q_{\alpha\beta}$ etc..
The first three terms are a bulk free energy density whose overall scale is set by $A_0$ (discussed further below); $\gamma$ is a control parameter, related to reduced temperature. Varying the latter in the absence of chiral terms ($q_0=0$) gives an isotropic-nematic transition at $\gamma = 2.7$ with a mean-field spinodal instability at $\gamma = 3$.

The rest of the free energy in Eq.\ref{free} describes distortions of the order parameter field. As is conventional \cite{blue1,deGennes} we assume that splay, bend and twist deformations of the director are equally costly; $K$ is then the one elastic constant that remains. The parameter $q_0$ is 
related via $q_0=2\pi/p_0$ to the pitch length, $p_0$, describing one full turn of the director in the cholesteric phase.
In BPs one observes that the pitch length (still defined locally by the spatial rotation rate of $\bf n$) slightly increases on entering the BP from the cholesteric phase. To account for this, a `redshift' factor $r$ is introduced \cite{yeomans2} whose variation effectively allows free adjustment of the BP lattice parameter, $\Lambda \to \Lambda/r$.  
To avoid changing the size of the simulation box, redshifting is performed in practice by an equivalent rescaling of the pitch parameter and elastic constant, $q_0\to q_0/r$ and $K\to K \,r^2$. In simulations aimed solely at free energy minimization, it is legitimate to make $r$ a dynamic parameter and update it on the fly to achieve this \cite{yeomans2}. We do not allow this during our domain growth runs, but do relax $r$ at the end of selected runs as part of our free energy comparison (see Section 6).

The phase diagram of blue phases (if thermal fluctuations can be neglected \cite{stark}) depends on just two dimensionless numbers,
which are commonly referred to as $\kappa$, the {\em chirality}, and 
$\tau$, the {\em reduced temperature} \cite{blue1}. 
In terms of the above parameters, these are:
\begin{eqnarray}\label{cntrl-param} 
\tau&=&\frac{27(1-\gamma/3)}{\gamma}\label{tau}\\
\kappa&=&\sqrt{\frac{108\ K\, q_0^2}{A_0\, \gamma}}\label{kappa}.
\end{eqnarray}
If the free energy density Eq.\ref{free} is made dimensionless, $\tau$ appears as prefactor of the term 
quadratic in ${\bf Q}$, whereas $\kappa$ quantifies the ratio between bulk and 
gradient free energy terms.

\subsection{Equations of motion}

A framework for the dynamics of liquid crystals is provided by the 
Beris-Edwards model \cite{beris}, in which the time evolution of the tensor order parameter obeys
\begin{equation}\label{eom}
\left(\partial_t+ u_\alpha \partial_\alpha \right){\bf Q} - 
{\bf S}({\bf W},{\bf Q}) = \Gamma {\bf H} + \mbox{\boldmath{$\zeta$}}
\end{equation}
where $\mbox{\boldmath{$\zeta$}}$ 
is a noise term discussed in section 4.
In the absence of flow, Eq.\ref{eom} describes 
relaxation towards equilibrium, with a rotational diffusion 
constant $\Gamma$, driven by the molecular field ${\bf H}$.
The latter is given by \cite{beris}
\begin{equation}
{\bf H}=-\frac{\delta {\cal F}}{\delta {\bf Q}}+ \frac{\bf I}{3} Tr\left(\frac{\delta {\cal F}}{\delta {\bf Q}}\right).
\end{equation}
The tensor ${\bf S}$ in Eq.\ref{eom} couples the order parameter
to the symmetric and antisymmetric parts of the velocity 
gradient tensor $W_{\alpha \beta}\equiv\partial_\beta u_\alpha$, defined as 
\begin{equation}
{\bf A}=({\bf W}+{\bf W}^T)/2; \; \; {\boldsymbol \Omega}=({\bf W}-{\bf W}^T)/2.
\end{equation}
This coupling term reads explicitly
\begin{eqnarray}\label{coupling-term}
{\bf S}({\bf W}, {\bf Q}) &=& (\xi {\bf A} + {\boldsymbol \Omega})({\bf Q}+\frac{\bf I}{3}) +({\bf Q}+\frac{\bf I}{3})(\xi {\bf A}- {\boldsymbol \Omega}) \nonumber\\ 
&-&2 \xi ({\bf Q}+\frac{\bf I}{3})Tr({\bf Q W}).
\end{eqnarray}
Here $\xi$ is a material-dependent `tumbling parameter' that controls the relative importance of rotational and elongational flow for molecular alignment. We choose $\xi = 0.7$, within the `flow aligning' regime for which molecules align at a fixed angle (the Leslie angle) to the flow direction in weak simple shear \cite{deGennes}. (For `flow tumbling' materials, the director instead rotates continuously \cite{deGennes,beris}.)

The momentum evolution obeys a Navier-Stokes equation driven by the divergence of a stress tensor ${\boldsymbol \Sigma}$:
\begin{equation}\label{nse}
\rho\,\partial_tu_\alpha +\rho \,u_\beta \partial_\beta u_\alpha=\partial_\beta {\Sigma}_{\alpha\beta}
\end{equation}
This pressure tensor is in general asymmetric and includes both viscous and thermodynamic components 
\begin{eqnarray}
\Sigma_{\alpha \beta}&=&-P_0 \delta_{\alpha \beta}  +  \eta \{ \partial_\alpha u_\beta + \partial_\beta u_\alpha\}\nonumber\\
&-&  \xi H_{\alpha \gamma}\left(Q_{\gamma \beta} +\frac{\delta_{\gamma \beta}}{3} \right) -\xi \left(Q_{\alpha \gamma} +\frac{1}{3} \delta_{\alpha \gamma}\right) H_{\gamma \beta}  \nonumber\\ 
&+& 2 \xi  \left(Q_{\alpha \beta} +\frac{1}{3} \delta_{\alpha \beta}\right) Q_{\gamma \nu} H_{\gamma \nu}-\partial_\alpha Q_{\gamma \nu} \frac{\delta{\cal F}}{\delta \partial_{\beta} Q_{\gamma \nu}}\nonumber\\
&+& Q_{\alpha \gamma}H_{\gamma \beta}-H_{\alpha \gamma} Q_{\gamma \beta}
\end{eqnarray}
Note that within the lattice Boltzmann flow solver, the isotropic pressure and viscous terms are managed directly by the solver (as in a simple Newtonian fluid, of viscosity $\eta$) whereas the divergence of the remaining terms is treated as a local force density on that fluid.

\subsection{Parameter mapping to physical units}

Here we describe how simulation parameters are
related to physical quantities in real BP materials.
In order to get from simulation to physical units, we need to
calibrate scales of mass, length, and time (or equivalently length, energy and time). We follow a methodology similar to that of \cite{yeomans1}.

First we define a set of LB units (LBU) in which the lattice parameter $\ell$, the time step $\Delta t$, and a reference fluid mass density $\rho_0$ are all set to unity. This is the set of units in which our algorithm is actually written. The first two of these parameters directly connect to observables in the simulations and we show below how to map these onto experiments. The fluid density, however, enters differently. So long as fluid inertia remains negligible (low Reynolds number, Re = $\rho V\Lambda/\eta$, where $V,\Lambda$ are typical length scales and velocities of any flow) the physics observed will not depend on the actual mass density $\rho$ of either the physical or the simulated system.  Since LB uses inertia to update the fluid velocities, it improves efficiency to use a density that causes Re to be several orders of magnitude larger than in experiments; so long as Re remains small (say, Re $< 0.1$), no harm is done \cite{codef}. Such parameter steering is helped by allowing $\rho\neq\rho_0$ within the code.

We now turn to the calibrations of length, energy and time. The length scale calibration is straightforward, and fixed by the cholesteric pitch $p_0$, which
is typically in the 100-500 nm range \cite{blue1}. More precisely, in our simulations we set the unit cell of BPI/II to be 16 LBU;  this gives good resolution without wasting resource. Therefore the LBU length unit (one lattice site) corresponds to, say, 10nm in physical space.

To get an energy scale, we use the measurements cited in Appendix D of \cite{blue1}, which suggest
\begin{equation}
\frac{a}{b^2} = \frac{27}{2 A_0\gamma} \sim 2-5 \times 10^{-6}
{\rm J}^{-1}\, {\rm m}^{3}.
\end{equation}
Here $a,b$ are parameters defined in \cite{blue1}, and the required ratio is expressed in terms of our chosen parameters (whose comparison with \cite{blue1} shows that $b\sqrt{6} = A_0\gamma/3$ and $a = A_0\gamma/4$). From this relation, given that $\gamma \simeq 3$, we obtain that $A_0 \simeq 10^6$ Pa. On the other hand, our simulations use a value of $A_0\simeq 0.01$ LBU. This requires that the LB unit of stress is about $10^8$ Pa in SI units.  We next use this to calibrate the LB time unit and then crosscheck that the resulting fluid density gives acceptable Reynolds numbers.

For the timescale calibration we use the
following formula \cite{deGennes,yeomans1} which relates the so-called `rotational viscosity' $\gamma_1$ (defined by the relaxation equation $\gamma_1\partial {\bf n}/\partial t = -\delta F/\delta {\bf n}$) for the director field in a well-aligned nematic, to the ordering strength $q$ and the order parameter mobility $\Gamma$:
\begin{equation}
\gamma_1=\frac{2q^2}{\Gamma}\;.
\end{equation}
In our simulations, we choose $\Gamma = 0.3$ and also select the thermodynamic parameters to give $q\simeq 0.3$. Therefore $\gamma_1 \simeq 0.6$ LBU.  
For real materials, $\gamma_1$ lies usually in range $10^{-2}-10^{-1}$ Pa s \cite{deGennes}; we choose for definiteness $\gamma_1 = 0.06$ Pa s  $= 0.6$ LBU. Given the previous result for stress, this requires that the LB time unit (one algorithmic timestep) equates to $10^{-9}$ s. 

Note that in our simulations we use a fluid viscosity $\eta \simeq 1-2$ LBU  $= 0.1-0.2$ Pa s (using the parameter mapping just established). This is sensible (if somewhat low) for a molecular nematogen in the isotropic phase. (The effective viscosity in ordered phases is of course higher, but the coupling to the order parameter handles this.)
Similarly, we adopt elastic constant values $K\simeq 0.01$ LBU $ = 10^{-10}$ N, corresponding to a Frank elastic constant $K/2$ of $50$ pN which is again sensible \cite{deGennes}. 

Finally, we need to cross-check the fluid density. We have 1 kg m$^{-3} = 1$ Pa s$^2$ m$^{-2} = 10^{-8+18-16}$ LBU $= 10^{-6}$ LBU. Thus the reference density $\rho_0$ equates to a fluid density $\rho = 10^6$ kg m$^{-3}$, roughly a thousand times larger than experimental values. As explained previously, however, this makes no difference so long as the Reynolds number Re $= \rho V\Lambda/\eta$ remains small enough. This dimensionless number can be evaluated directly in lattice units. We have $\rho$ and $\eta$ of order unity, and set $\Lambda \sim 16$ (a BP unit cell). Observing that typical velocities arising in our simulations are around $10^{-5}$ LBU we get Re $\lesssim 10^{-3}$. Even allowing for higher peak velocities and larger $\Lambda$ in some materials, this is safely small \cite{codef}. 

To summarize the above, our simulations faithfully represent experimentally realisable BP-forming materials, subject to the interpretation of the LB units for length, time, and energy density are close to 10 nm, 1 ns, and 100 MPa respectively.

\subsection{Adding noise}

With these choices, the typical difference in free energy density between say
BPI and BPII is $10^{-6}-10^{-5}$ LBU (see Table S2) or about 
$100-1000$ Pa. Even for a rather small BP lattice constant of $160$ nm, this
corresponds to a free energy difference per unit cell of $100-1000$ $k_BT$,
much larger than any thermal energy. Such unit-cell level differences represent a reasonable estimate of the barrier to topological reconnections within an evolving BP structure, and appear to preclude any significant role for thermal noise in the domain growth process.
Nonetheless it has been argued theoretically \cite{stark} that entropy plays an important role in BP thermodynamics and presumably also therefore, domain growth. (This certainly becomes more likely for atypically small $A_0$ and/or small unit cells.) We therefore repeated selected simulations with thermal noise present.

We choose to add noise to the order parameter sector only, via the $\zeta_{\alpha\beta}$ term in Eq.\ref{eom}. In principle, noise can also be added to the fluid mechanical sector \cite{rjoy}; this will create an additional conserved diffusion of the order parameter (a kind of Brownian motion) which should however be negligible, at least at large length scales, compared to the local nonconservative relaxation embodied in Eq.\ref{eom}.
The fluctuation-dissipation theorem then fixes the variance of the noise in Eq.\ref{eom} as
\begin{equation}
\langle \zeta_{\alpha\beta}({\bf r},t)\zeta_{\gamma\delta}({\bf r'},t')\rangle
= 2 k_BT\Gamma P_{\alpha\beta\gamma\delta}\delta({\bf r}-{\bf r'})\delta(t-t').
\end{equation}
Here $\bf P$ is a tensor that projects the order parameter into independent components (respecting its symmetry and tracelessness) and allows only diagonal correlations in that space. 

By using the parameter mapping detailed above, we can find the value of 
$k_BT = 4\times 10^{-21}$J in simulation units as $4\times 10^{-5}$ LBU. 
However, to allow for variation in material parameters, such as the BP unit cell size, we have studied a range of noise temperatures in the range $10^{-6}$--$10^{-4}$ LBU. At the upper end of this range we can observe vibrant thermal fluctuations of the disclination network, associated with some shifting of phase boundaries. This merits further study in view of the claims of \cite{stark} that fluctuations can signigicantly alter the equilibrium physics of blue phases, although a full exploration of such effects lie beyond the scope of the current work. However, in terms of domain growth dynamics, we have found the main effects of adding modest amounts of thermal noise to be quantitative rather than qualitative. (Here `modest' means $k_BT \le 10^{-5}$ LBU, as might be relevant to BPs with lattice parameters $\ge 300$ nm rather than the 160nm value used in the parameter mapping in section 3.) An exception was already discussed in relation to Fig.S1, which shows that quite low noise levels can disrupt the orderly but aperiodic growth that would otherwise arise from a rod-like nucleus. 

\subsection{Quench details}

Quenches were performed as described in the Methods section of the main text. Supporting Table S1 specifies the complete simulation parameters (all in LBU) for each of the runs presented in Figures 2-5 of the main text and  Figures S1-S2.

\subsection{Free energy comparisons}

Supporting Table S2 compares the free energy densities of three crystalline blue phases (BPI, BPII, O5) with the cholesteric phase and the end-state amorphous networks for the final $\tau,\kappa$ values appropriate to all runs reported in Figs.2-5 and Figs.S1,S2. (The noise-free isotropic phase has free energy density zero in all cases.) Note that O5, as long predicted theoretically \cite{blue1}, becomes the most stable ordered phase at high chirality. However this phase is not seen experimentally and its relative stability might be a consequence of the one elastic constant approximation or some other shortcoming of the free energy, Eq.\ref{free}. 

For end-state amorphs nucleated within an isotropic phase, two free energy values are reported. The higher one is the direct result of the quench, performed at fixed redshift (as is appropriate when simulating directly the equations of motion presented above). The lower value is for a annealing protocol whereby the redshift is released at the end of the main part of the simulation, either when the disclination network has filled the simulation box and rearrangements have come to a virtual standstill (this was done for $\tau =0, \kappa = 2$) or when the disclination network first collides with its own periodic images (this was done for $\tau =0, \kappa =3$). This `devil's advocate' annealing schedule finds the best possible free energy among states with topology close to the amorph, regardless of whether such states were dynamically accessible from the initial nucleus. 

For $\kappa \le 2$, the final redshift release makes no qualitative difference to the observed topology; moreover the free energies found after this procedure for ($\kappa = 2$) still lie above those of the ordered `target' structure, as they do for all runs with $\kappa \le 2$ done in the absence of a final redshift-release step. This confirms that the amorphous end-state is metastable, as claimed in the main text. 

The quenches at $\tau = 0, \kappa = 3$ offer a test of Ostwald's rule of stages; for these parameters, the free energy of the final amorphous network (with or without redshift release, the latter in bold) lie further in free energy from the initial isotropic phase ($f=0$) than the structure that was nucleated. When this is BPI, so do all remaining ordered phases. Thus the nucleus is now made of the phase whose free energy lies closest below that of the initial bulk; by Ostwald's rule, this should be the first phase to grow. Despite this, an amorphous network is formed, and the initial nucleus disappears -- just as it did when the nucleated phase was the stable one. Therefore our results have no explanation in terms of Ostwald's rule; indeed, for this system, they disprove it.

In these high chirality runs, the final release of the redshift noticably changes the dynamics, accelerating the continuous ripening of the amorphous network towards the equilibrium O5 phase. No redshift release was performed for runs involving nuclei within a cholesteric matrix ($\tau = -0.5, \kappa = 1.2,1.35$)  due to the much longer simulation times required in this case. Indeed, for runs (F2,F3) the defect network did not fill the box by the end of the run and the quoted free energy densities in these cases are upper bounds. However, correcting for such volume-fraction effects shifts the values for the amorphs down by less than one percent, confirming their metastability.

The amorphous networks we report in this work appear closely related to BPIII, an equilibrium blue phase, believed to be amorphous, which is found experimentally at high chiralities. We address elsewhere (manuscript in preparation) the question of whether the chosen free energy density, Eq.\ref{free}, can indeed predict a stable BPIII phase in that regime. Of significance here is the fact that it does predict a {\em metastable} BPIII phase for $\tau = 0$ and $\kappa = 2,3$ (relevant to Figs.4,5, and S2). To establish this, we have created candidate BPIII structures by evolving an initial state consisting of randomly oriented and positioned double twist cylinders (DTCs), embedded in a cholesteric matrix. On suitable annealing (with redshift enabled), these relax to form amorphous networks, metastable relative to BPI/II, that on visual inspection look very similar to our end-state structures. However, for $\tau =0, \kappa = 2,3$, the DTC-based networks have a lower free energy ($\tilde f = 10^5f = -3.104, -1.144$) than the amorphs formed in our nucleation runs with BPI/II. (Except for the case of the BPII-nucleated amorph at $\kappa = 3$, this remains true even if we violate the true dynamics by releasing the redshift.) Thus our end-state amorphs are {\em not} the disordered network phase of lowest free energy in any of these cases. Arguably however BPIII {\em is} that phase: certainly it must be so at higher chirality, where it is thermodynamically stable. Thus we do not designate our amorphs as belonging directly to a metastable branch of BPIII.

\begin{table*}[ht]
{\renewcommand{\arraystretch}{1.0}
\renewcommand{\tabcolsep}{0.1cm}
\begin{tabular}{|c|c|c|c|c|c|c|c|c|c|c|c|}
\hline
Fig. & $\tau_{eq}$ & $\kappa_{eq}$ & $A_{0, eq}$& $\gamma_{eq}$ & $\tau$ & $\kappa$ &$A_0$ & $\gamma$ & $r$ & $K$ &  $k_B T$ \\
\hline 
\hline
2 & 0.75 & 2& 0.0075 & 2.769 & 0 & 2& 0.0069 & 3 & 0.91& 0.02& 0 \\
3 & -0.5 & 0.45 & 0.1295 & 3.176 & -0.5 & 1.2 & 0.0182 & 3.176 & 0.83& 0.01& 0\\
4 & -0.5 & 0.45 &0.1295 & 3.176& -0.5 & 1.35& 0.0144 &  3.176 & 0.83& 0.01& 0\\
S1 & 0.95 & 0.4 & 0.1918 & 2.714 & 0 & 2& 0.0069  & 3 & 0.83& 0.01& 0\\
S2 & -0.5 & 0.45& 0.1295 & 3.176 & -0.5 & 1.35 & 0.0144 &  3.176 & 0.83& 0.01 &$8.33\times 10^{-6}$ \\
S3 & 0.95 & 0.4 & 0.1918 & 2.714 & 0 & 3 & 0.0031 &  3 & 0.83& 0.01 & 0\\
\hline
\end{tabular}
}
\caption{[Table S1] Simulation parameters in lattice units: temperature $\tau_{eq}$, chirality  $\kappa_{eq}$, bulk free energy constant $A_{0, eq}$ and $\gamma_{eq}$ at the equilibration point, temperature $\tau$, chirality $\kappa$, bulk free energy constant $A_0$ and $\gamma$ after the quench, redshift $r$ as used during the equilibration and the main part of the simulation, elastic constant $K$,
and noise temperature $k_BT$. For all runs the  pitch $q_0 = \pi/16$; rotational diffusion constant $\Gamma = 0.3$; the tumbling parameter $\xi=0.7$; the viscosity $\eta = 5/3$, and the fluid density $\rho = 2$.} 
\end{table*}

\begin{table*}[h]
{\renewcommand{\arraystretch}{1.0}
\renewcommand{\tabcolsep}{0.1cm}
\begin{tabular}{|c|c|c|c|c|c|c|c|}
\hline
State & Fig. & $\tau$ & $\kappa$ & $r$ & $\tilde f(r)$ & $r'$ & $\tilde f(r')$\\
\hline 
\hline
BPI-ISO-D & S4 & 0 & 2& 0.83 & -2.949 & 0.946 & -3.066\\
BPII-ISO-D & 2 & 0 & 2& 0.91 &-3.048 &0.972 & -3.088\\
CH & - & 0 & 2&- &- &1.006 &-0.826 \\
BPI & - & 0 & 2&- &- &0.846 &-2.780 \\
BPII & - & 0&2 &- &- &0.908 &{\em -3.162} \\
O5 & - & 0& 2&- &- & 1.009  &-3.048 \\
\hline
BPI-ISO-D & S3 & 0 & 3 &0.83 &{\bf -0.938} &0.949 & { -1.043}\\
BPII-ISO-D & - & 0 & 3 &0.91 & {\bf -1.135} &0.959 & { -1.149}\\
CH & - &0 &3 &- &- &1.006 & 0.003\\
BPI & - &0 &3 &- &- &0.921 &-0.095 \\
BPII & - &0 &3 &- &- &0.919 &-1.063 \\
O5 & - &0 &3 &- &- &1.020 &{\em -1.192} \\
\hline

BPI-CH-D & 3 & -0.5 & 1.2 & 0.83 &-17.749 &- &- \\
CH & - &-0.5 &1.2 &- &- &1.006 &-16.466 \\
BPI & - &-0.5 &1.2 &- &- &0.819 &{\em -19.231} \\
BPII & - &-0.5 &1.2 &- &- &0.877 & -18.973 \\
O5 & - &-0.5 &1.2 &- &- &0.952 &-17.949 \\
\hline
BPI-CH-R & 4 & -0.5 & 1.35 & 0.83 &-13.580 &- &- \\
BPI-CH-RN & S2 & -0.5 & 1.35 & 0.83 &-13.002 &- &- \\
CH & - &-0.5 &1.35 &- &- &1.006 & -11.825\\
BPI & - &-0.5 &1.35 &- &- &0.826 &{\em -14.720} \\
BPII & - &-0.5 &1.35 &- &- &0.887 & -14.615 \\
O5 & - & -0.5&1.35 &- &- &0.969 &-13.743 \\
\hline
\end{tabular}
}
\caption{[Table S2] Free energy densities $\tilde f(r) \equiv 10^5 f(r)$ in LBU for the ordered blue phases (BPI, BPII and O5) and the cholesteric phase (CH) and for the amorphous networks grown from a BPI-droplet in isotropic (BPI-ISO-D) or cholesteric environment (BPI-CH-D), a BPII-droplet in isotropic environment (BPII-ISO-D) or from a BPI-rod in cholesteric environment with (BPI-CH-RN) and without noise (BPI-CH-R). The free energy density $f(r)$ of each phase was attained by direct evolution of the initial state at fixed redshift $r$; $f(r'),r'$ are corresponding values after redshift release. Numbers in bold provide tests of Ostwald's rule(see supplementary Text S1, section 6); italics identify the equilibrium phase at given $\tau,\kappa$.} 
\end{table*}

\begin{figure*}[ht]
\includegraphics[width=0.5\columnwidth]{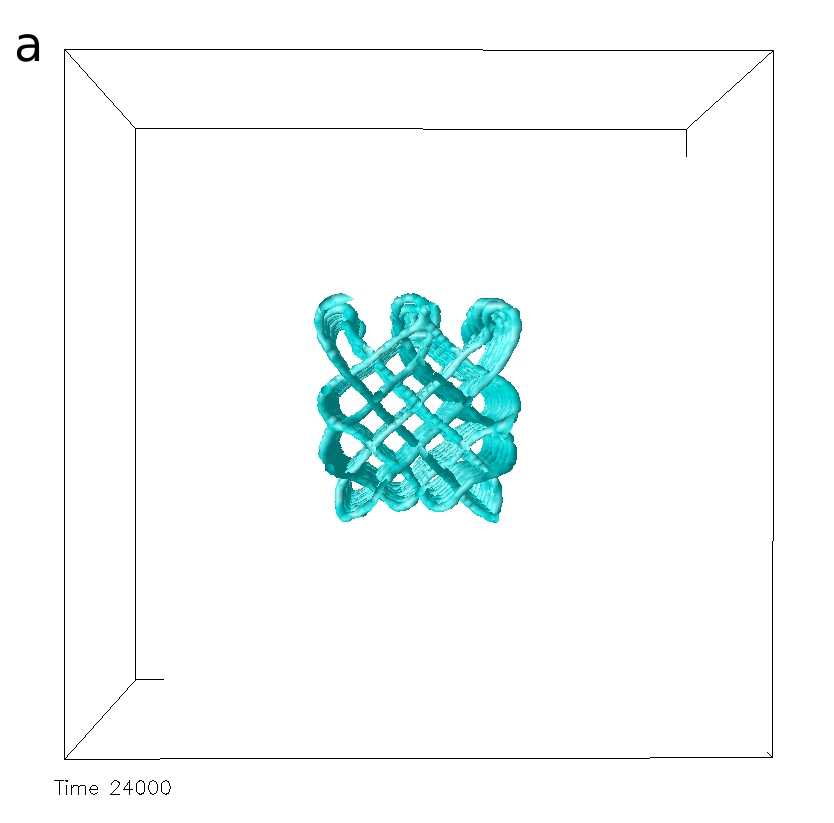}
\includegraphics[width=0.5\columnwidth]{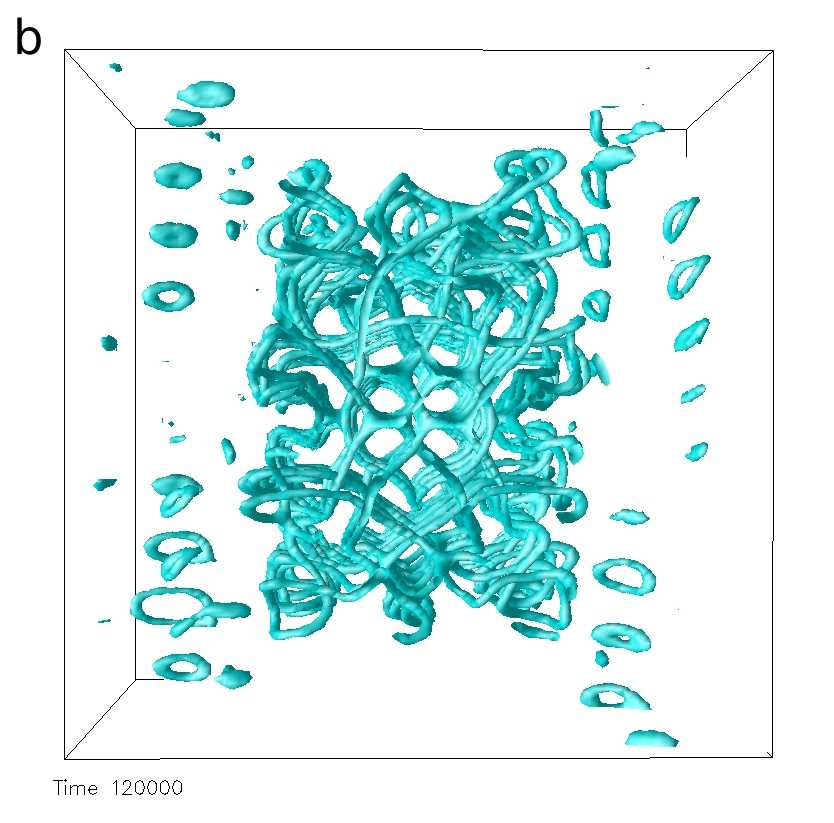}
\includegraphics[width=0.5\columnwidth] {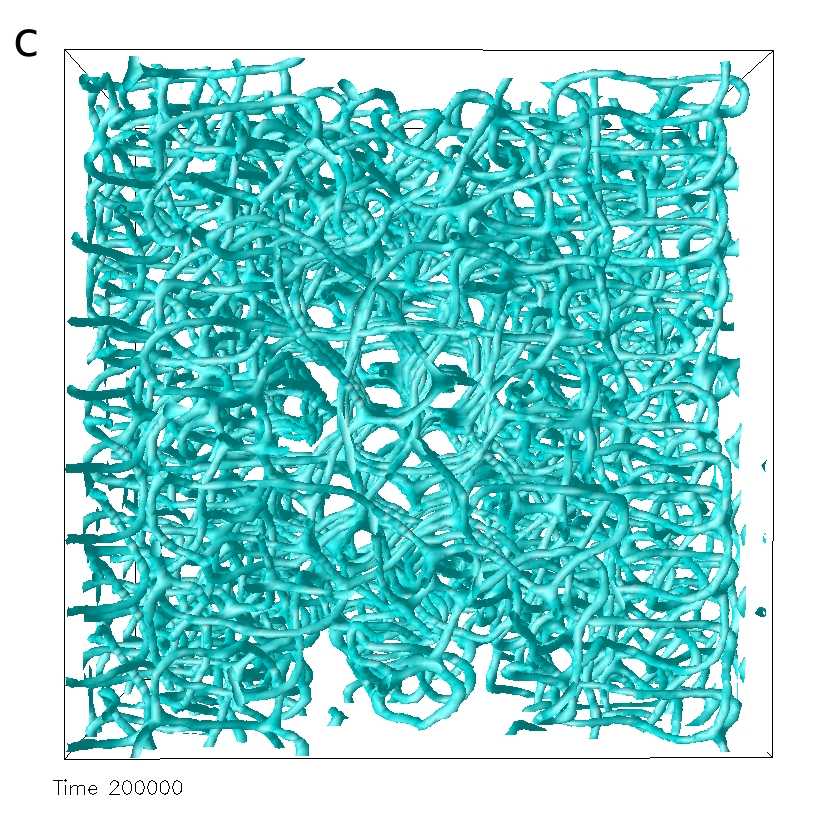}
\includegraphics[width=0.5\columnwidth]{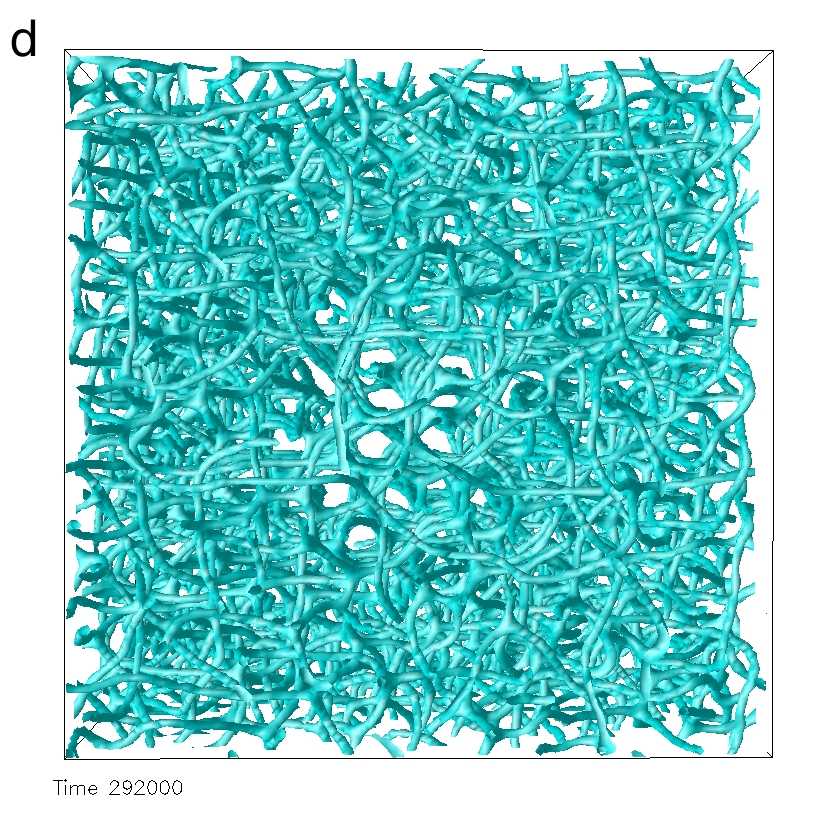}
\caption{[Fig.S1] Domain growth of a BPI rod in cholesteric environment including noise in the order parameter equation of motion. The noise strength was set to $k_B T=8.33\cdot 10^{-6}$ LBU. The pictures show isosurfaces of the order parameter at $t=2.4\times 10^4$ timesteps (a); $t=1.2\times 10^5$ (b); $t= 2\times 10^5$ (c); and $t=2.9\times 10^5$ (d). The equilibration and quench were performed in exactly the same way as in the simulation of the BPI rod without noise (Figure 4 in main text). Note that from about $t=10^5$ onwards, the disclination network ceases loses the symmetry of the initial state.}
\end{figure*}

\begin{figure*}[ht]
\includegraphics[width=0.5\columnwidth]{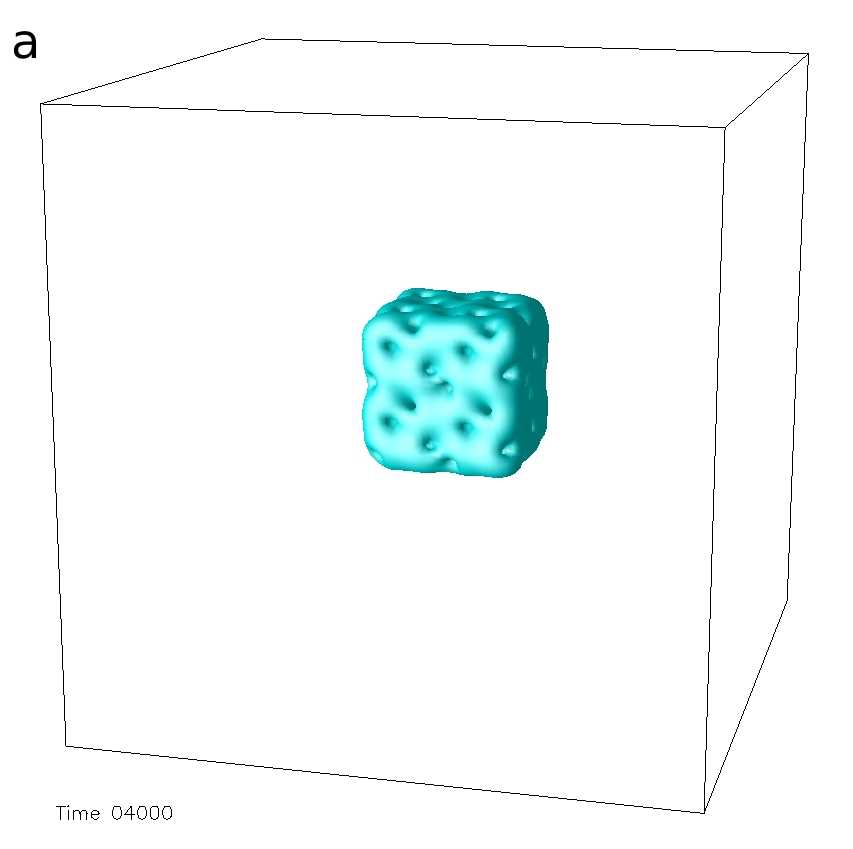}
\includegraphics[width=0.5\columnwidth]{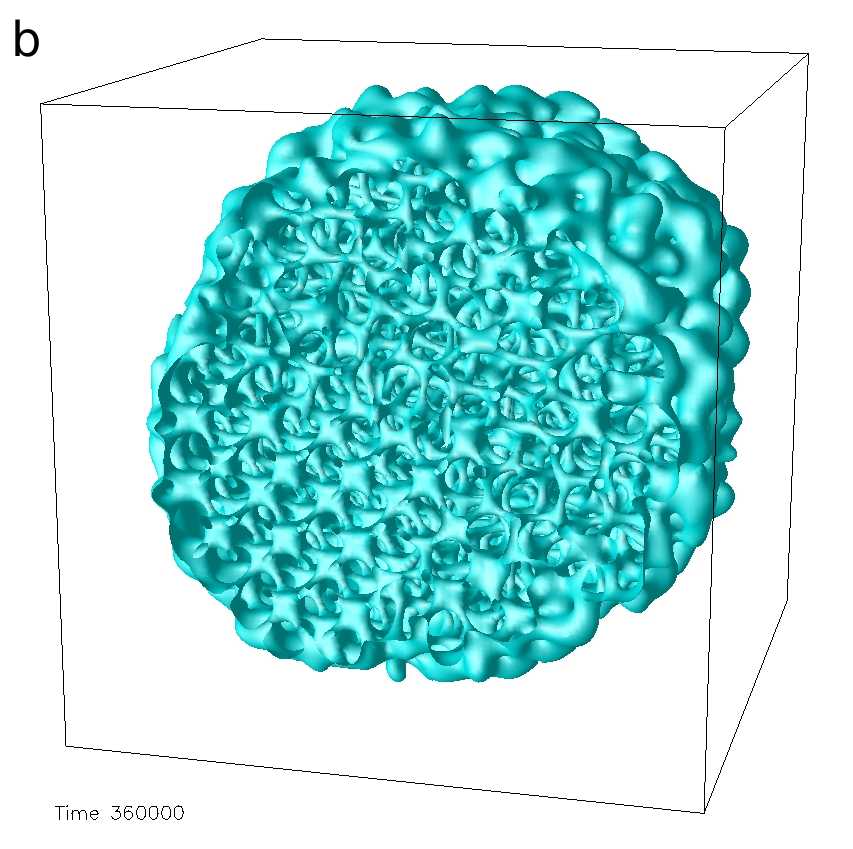}
\includegraphics[width=0.5\columnwidth]{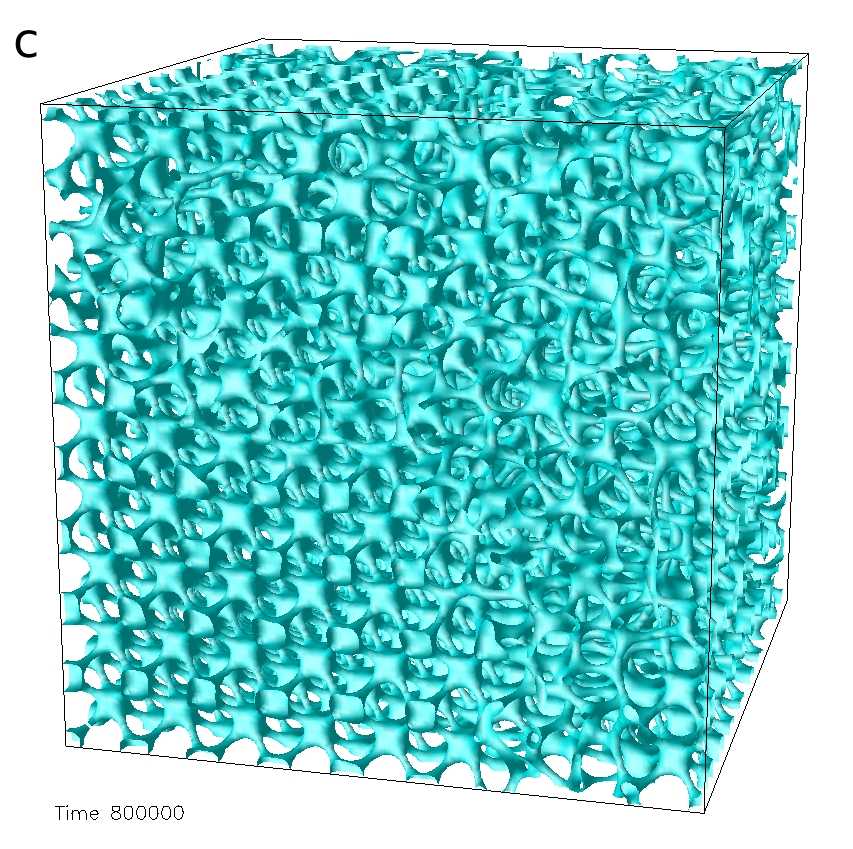}
\includegraphics[width=0.5\columnwidth]{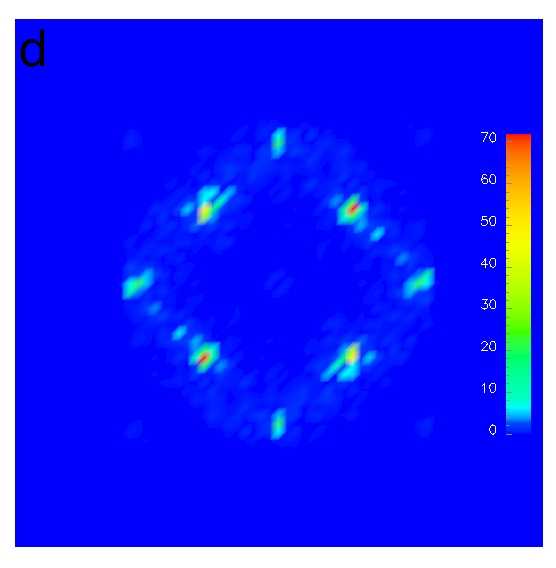}
\caption{[Fig.S2] Domain growth of a BPI droplet in isotropic environment at high chirality. The droplet was equilibrated near the isotropic-BPI boundary at $\tau=0.95, \kappa=0.4$ and then quenched to $\tau=0, \kappa=3$. At this point O5 is the thermodynamically stable phase. The pictures show the isosurface ($q=0.12$) of the scalar order parameter during the equilibration (a), at an intermediate stage (b) and in the final state of the simulation (c). In (b) the isosurface has been cut at $y=32$ to reveal the internal structure. In (d) a cut through the structure factor of the final state (c) along $k_z=0$ for $k_x\le k_y$ on the interval $[-\pi/2\ell,\pi/2\ell]$ is depicted. Although the Bragg-peak pattern appears rather blurred, the cubic signature of O5 is clearly recognizable.}
\end{figure*}

\end{document}